\documentclass[showpacs,aps,prx,twocolumn,superscriptaddress,longbibliography]{revtex4-2}

\usepackage[english]{babel}
\usepackage[T1]{fontenc}
\usepackage[utf8]{inputenc}
\usepackage{amsmath}
\usepackage{amsthm}
\usepackage{amsfonts}
\usepackage{xcolor}
\usepackage{xr-hyper}
\usepackage{hyperref}
\usepackage{amssymb,graphicx}
\usepackage{tikz}
\usetikzlibrary{decorations.pathmorphing}
\usepackage{siunitx}
\usepackage{mathbbol}
\usepackage{braket}
\usepackage{verbatim}
\usepackage{tabularx}
\usepackage{qcircuit}
\usepackage{stmaryrd}
\newcommand{\Secref}[1]{Sec.~\ref{#1}}
\newcommand{\Figref}[1]{Fig.~\ref{#1}}
\newcommand{\Eqref}[1]{Eq.~(\ref{#1})}

\newsavebox\MBox

\newcommand*{\colorboxed}{}
\def\colorboxed#1#{%
  \colorboxedAux{#1}%
}
\newcommand*{\colorboxedAux}[3]{%
  \begingroup
    \colorlet{cb@saved}{.}%
    \color#1{#2}%
    \boxed{%
      \color{cb@saved}%
      #3%
    }%
  \endgroup
}

\begin{document}

\title{Simulation of quantum annealing on a semiconducting cQED device \\ for Multiple Hypothesis Tracking (MHT) benchmark}

\author{Q.~Schaeverbeke }
\thanks{These authors contributed equally to the development of the quantum annealing framework and its application.}
\affiliation{C12 Quantum Electronics, Paris, France}

\author{V.~Radovi\'{c} }
\thanks{These authors contributed equally to the development of the quantum annealing framework and its application.}
\affiliation{C12 Quantum Electronics, Paris, France}

\author{JM. Divanon }
\thanks{These authors provided the classical algorithm for path estimation.}
\affiliation{Thales Research \& Technology, Singapore}

\author{Bing Hong Teh }
\affiliation{Thales Research \& Technology, Singapore}
\thanks{These authors provided the classical algorithm for path estimation.}

\date{\today}

\begin{abstract}
    We explore the expected performance of a semiconducting spin cQED quantum processor for Multiple Hypothesis Tracking (MHT) algorithm via a quantum annealing procedure.
    From two different benchmarking scenarios we evaluate this type of quantum annealer on a quantum emulator in which we incorporated both dynamical coherent errors and incoherent errors.
    From estimate of the reset, measurement and annealing time of the processor, we find that cQED-spin processors could reach a total run time of around \SI{50}{\milli\second}.
    This makes this technology promising for potential real time application such as radar tracking.
\end{abstract}

\maketitle

\section{Introduction}
Semiconducting spin qubits are a promising technology for quantum computing \cite{chatterjee2021semiconductor,burkard2023semiconductor}.
They have exhibited long coherence times, on the order of several \si{\milli\second} \cite{veldhorst2014addressable} and fast operating times, with single and two qubit gate times under the micro second, reaching average gate fidelities above $99\%$ for two qubit gates and close to $99.999\%$ for single qubit gates \cite{veldhorst2015two,takeda2016fault,noiri2022fast,xue2022quantum,dijkema2025cavity,wu2025simultaneous}.
Taking advantage of their small physical imprint and an already well mastered industry, they are often presented extremely promising for scaling processors to a large number of qubits.
A general limitation of the semiconducting spin qubits is their low connectivity, however this can be mediated through their integration in circuit quantum electro-dynamics (cQED) architectures \cite{dijkema2025cavity}.
Indeed, flopping-mode qubits are essentially artificial molecules that present a large dipole which can be efficiently coupled to a microwave resonator.
In this way, multiple qubits can couple through the same resonator, unleashing long range interactions between them.
Semiconducting spins integrated in cQED architecture have a long way to go as, in particular, their typical coherence times are not on pare with other semiconducting devices, reaching only a few \si{\micro\second} \cite{neukelmance2025microsecond}, but they already demonstrated the ability to perform two qubit gates \cite{dijkema2025cavity}.
In this article, we propose to explore the potential of semiconducting spin in high connectivity platforms for practical applications.
More specifically, we simulate the benchmark of a quantum annealer implemented on a semiconducting spin cQED processor on a Multiple Hypothesis Tracking (MHT) algorithm.
Multi-target tracking has many practical use cases including air traffic control and surveillance, oceanography, autonomous vehicles and robotics, remote sensing, computer vision, and biomedical research and has a long lasting history dating from the 1960's \cite{mallick2013introduction}.
One of the major challenges in radar multi-target tracking is the Data Association Problem (DAP), which consists of assigning radar detection measurements to target tracks such that each measurement is associated with a most one target and each target receives at most one measurement per scan. This process is complexified by the presence of spurious measurements (false alarms), noise-contaminated measurements, target maneuvers, missed detections and the requirement to operate in real time.
DAP has been identified as NP-hard, since the number of possible associations grows exponentially over time \cite{oh2009markov}.
One algorithm proposed to tackle these challenges is the Multiple Hypothesis Tracking (MHT) proposed in 1976 by D. Reid \cite{reid2003algorithm}.
It was proposed in \cite{RoussetRouard2025benchmark} to use a quantum algorithm to perform a subroutine of the algorithm called the Maximum Weighted Independent Set (MWIS).
In this article we explore the performance of a semiconducting spin cQED processor used as a quantum annealer for implementing a multi-target tracker.
We develop a noise model to account for the limits of the processor.
The noise model is ported into a quantum emulator, namely \textit{Callisto} developed at \textit{C12 Quantum Electronics}, that is used to estimate the feasibility of using \textit{C12}'s hardware for multi-target tracking.
The emulator originally framed as a gate based model, can simulate up to 20 qubits.
For the needs of this project, we developed an annealing feature for the emulator that deviates from the gate based approach.
We then estimate the total required running time and find that it would be consistent with a real time use as it is of the order of \SI{10}{\milli\second} when parallel qubit measurement is enabled.
In \Secref{S1} we present the algorithm that is tested on the emulator then in \Secref{S2} we present the hardware and how it is used for quantum annealing.
We briefly present the noise model and in \Secref{S3}, explain how the emulator is able to model the system.
Finally in \Secref{S4}, we present our results and conclude.

\section{Algorithm}\label{S1}
\subsection{MWIS algorithm}
The algorithm selected to harness DAP is the MHT algorithm introduced by Reid \cite{reid2003algorithm}.
It amounts to exploring a decision tree structure that represents every possible assignments.
For each node in the tree, a likelihood is computed.
Since the algorithm requires to keep track of the all measurement history, it is memory intensive.
With each new scan the path is updated as well as its likelihood, and the most likely path may change after each new scan.
More details on the MHT algorithm can be found in previous works \cite{RoussetRouard2025benchmark}.
One subroutine of the MHT algorithm involves solving the Maximum Weighted Independent Set (MWIS) problem.
This can be viewed as an undirected graph in which each vertex is a track hypothesis with a weight that corresponds to its likelihood and each vertices that are conflicting, meaning that they share a common measurement, are connected to each other.
The goal here is to select a set of independent vertices with the highest total weight.
Given a graph $G(V,E)$, the formulation of the problem is
\begin{equation}
    \label{mwis}
    \max_{x\in\{0,1\}}{\sum_{i\in V}w_i x_i}\; \text{ and }\; \forall (i,j)\in E, x_i+x_j\leq 1
\end{equation}
where $w_i$ is the weight of the vertex $i$ and $x_i=1$ if the hypothesis is selected or $x_i=0$ if not.
Considering the large size of the graph typically involved in real cases, it was proposed in \cite{RoussetRouard2025benchmark} that this routine was performed on a quantum computer using quantum annealing.
\subsection{Quantum annealing}
Quantum annealing is an optimization process for finding the global minimum of a function \cite{mohseni2022ising,rajak2023quantum}.
The function to minimize is encoded in an Ising Hamiltonian for which we are looking for the ground state.
Many optimization problems have been shown to be reducible to an Ising Hamiltonian, including the traveling salesman problem, the job scheduling problem, or the knapsack problem.
This shows that quantum annealing goes beyond strictly physical problems.
Quantum annealing consists in starting from a Hamiltonian with a trivial ground state, in general this is referred to as the driver Hamiltonian and adiabatically evolving the system toward the target Hamiltonian.
Following an adiabatic process, the trivial ground state will then evolves toward the target ground state and therefore the solution of the optimization problem.
The annealing process is summarized in the evolution equation:
\begin{equation}
    H=f(t)H_d+h(t)H_t,
\end{equation}
where $f$ and $h$ are the scheduling functions, such that $f(0)\gg h(0)$ and $f(t_f)\ll h(t_f)$, $t_f$ being the final time of the evolution.
$H_d$ is the driver Hamiltonian, in this paper it corresponds to the qubits initialized in their respective ground states:
\begin{equation}
    H_d=\frac{\omega_0}{2}\sum_k\sigma_z^{(k)},
\end{equation}
where $\sigma_i^{(k)}$ for $i\in\{x,y,z\}$ are the Pauli matrices of the qubit $k$.
Finally, $H_t$ is the target Hamiltonian that takes the form
\begin{equation}
    H_t=\sum_{i\neq j}\frac{J}{2}\sigma_x^{(i)}\sigma_x^{(j)}+\sum_k\Omega_k\sigma_x^{(k)}.
\end{equation}
\section{cQED semiconducting device}\label{S2}
\begin{figure}
    \centering
    \resizebox{\columnwidth}{!}{%
    \begin{tikzpicture}[scale=0.5]
        \node at (0,0) {\includegraphics[width=\columnwidth]{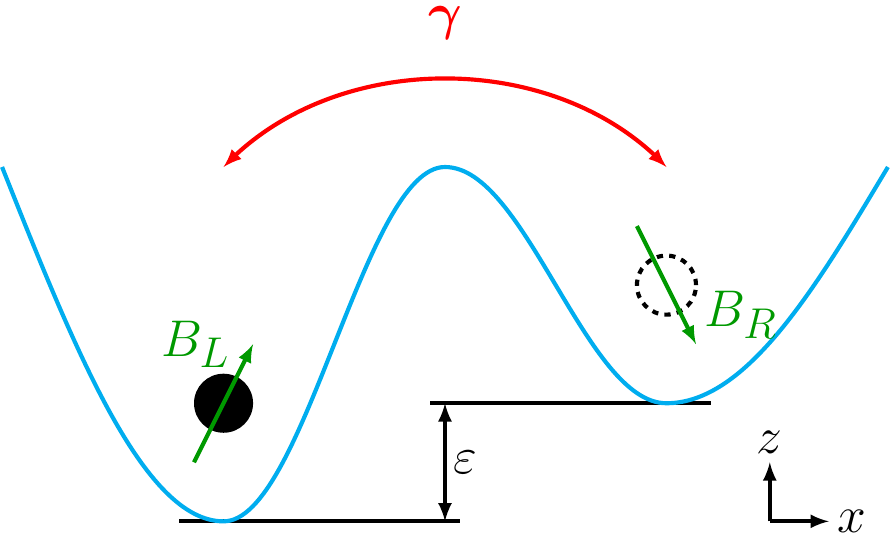}};
        \node at (0,-12) {\includegraphics[width=\columnwidth]{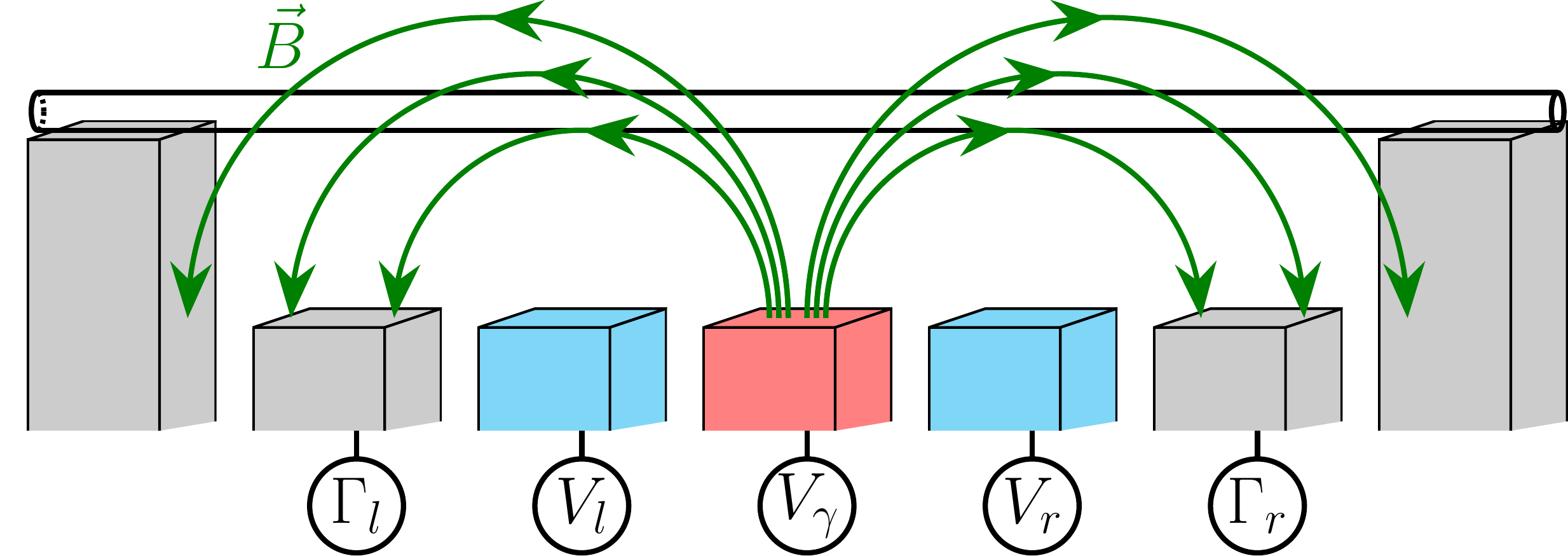}};
        \node at (-8,5) {\textbf{(a)}};
        \node at (-8,-7) {\textbf{(b)}};
    \end{tikzpicture}}
    \caption{Schematic of the carbon nanotube spin qubit. In panel \textbf{(a)}, a single electron trapped in a double potential well (blue curve). The detuning between the left and right wells is $\varepsilon$ while the tunneling rate between the dots is $\gamma$. An in-homogeneous static magnetic field is applied ($B_\alpha$ on dot $\alpha$). Panels \textbf{(b)} shows an experimental layout of the qubit where a nano-magnet burried under the red gate generates the spin texture. The red electrode controls the tunneling rate $\gamma$, while the blue electrodes control the detuning $\varepsilon$. $\Gamma_l$ and $\Gamma_r$ control the source and drain tunneling rates with the DQD.
    }
%}
    \label{f1}
\end{figure}
\begin{figure}
    \centering
    \includegraphics[width=\columnwidth]{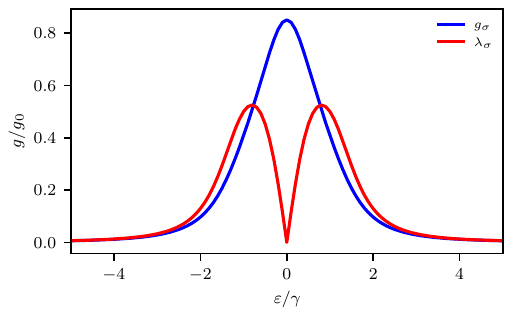}
    \caption{Logitudinal, $\lambda_\sigma$, and transversal, $g_\sigma$, spin-photon coupling as a function of the bias DQD energy $\varepsilon$. The bias energy is given in units of the tunneling energy $\gamma$, while coupling energies are given in unit of the bare electron-photon coupling $g_0$.}
    \label{f2}
\end{figure}
%

%
\begin{comment}
\begin{figure}
    \centering
    \begin{tikzpicture}

        \draw[color=cyan, line width=2pt] ({sqrt(2)},{-sqrt(2)})--(0,-2);
        \draw[color=cyan, line width=2pt] ({sqrt(2)},{-sqrt(2)})--(2,0);
        \draw[color=cyan, line width=2pt] ({sqrt(2)},{sqrt(2)})--(2,0);
        
        \draw[color=blue, line width=2pt] (0,2)--(0,-2);
        \draw[color=blue, line width=2pt] (0,2)--(-{sqrt(2)},-{sqrt(2)});
        \draw[color=blue, line width=2pt] (0,2)--(2,0);
    
        \fill[black] (0,2)circle(0.1);
        \fill[black] ({sqrt(2)},{sqrt(2)})circle(0.1);
        \fill[black] ({sqrt(2)},{-sqrt(2)})circle(0.1);
        \fill[black] ({-sqrt(2)},{-sqrt(2)})circle(0.1);
        \fill[black] ({-sqrt(2)},{sqrt(2)})circle(0.1);
        \fill[black] (-2,0)circle(0.1);
        \fill[black] (2,0)circle(0.1);
        \fill[black] (0,-2)circle(0.1);
        
    \end{tikzpicture}
    \caption{Schematic of the qubit connectivity for a given graph. Black circles represent qubits and blue lines represent a connection between two qubits. The darker the line the stronger the interaction.}
    \label{f3}
\end{figure}
\end{comment}
%

%
The device considered in this article is a semiconducting spin qubit embedded in a cQED architecture \cite{neukelmance2025microsecond}.
The spin qubit is made of a single electron double quantum dot (DQD) that acts as an artificial molecule.
\Figref{f1} shows a depiction of the spin qubit achieved in a carbon nanotube similarly to silicon or germanium spin qubits.
The DQD is achieved thanks to some dc gates which potential can be controlled.
A nano-magnet \cite{legrand2023} or ferromagnetic contacts \cite{neukelmance2025microsecond} are used to impose a spin orientation that differs in the two dots.
Finally one of the two dots is connected to a microwave resonator such that the qubits can be coupled two each other through a dispersive spin-spin interaction.
This corresponds to one blue gate in \Figref{f1}, being connected to a microwave circuit resonator.
The Hamiltonian describing a single qubit is
\begin{equation}
    H=\frac{\varepsilon}{2}\tau_z+\gamma\tau_x+\frac{\alpha_s}{2}\sigma_z+\frac{\alpha_{as}}{2}\sigma_x\tau_z+\omega_ca^\dagger a+g\left(a+a^\dagger\right)\tau_z,
\end{equation}
where $\varepsilon$ is the DQD bias energy, $\gamma$ is the DQD tunnel energy, $\alpha_{s}$ and $\alpha_{as}$ are the symetric and anti-symetric components of the magnetic potential energy, $\tau_i$ for $i\in\{x,y,z\}$ are the DQD orbitals Pauli matrices and $\sigma_i$ for $i\in\{x,y,z\}$ are the electron spin Pauli matrices.
The qubit is controlled during the annealing process through gates 2, 3 and 4, numbered from left to write in \Figref{f1} panel (b).
Gates 2 and 4 control the DQD bias energy while gate 3 controls the tunneling energy.
While the orbital-spin representation is practical to see directly the parameters upon which the experimentalists act directly, it does not show explicitly the encoding of the information.
In the computational basis of $H_q$ this encoding becomes more obvious and the Hamiltonian reads,
\begin{equation}\label{sh2}
    H=\frac{\omega_q}{2}\sigma_z+\omega_c a^\dagger a+(a+a^\dagger)H_D.
\end{equation}
Here $\sigma_i$ for $i\in\{x,y,z\}$ are not the spin Pauli matrices but the qubit Pauli matrices.
$\omega_q$, the qubit frequency is a function of $\varepsilon$, $\gamma$, $\alpha_{as}$ and $\alpha_s$ and $H_D$ is the ``dipole'' Hamiltonian that also depends on those parameters and represent the way the qubit couples to the resonator.
$H_D$ is comprised of two terms,
\begin{equation}
    H_D=\lambda_\sigma\sigma_z+g_\sigma\sigma_x,
\end{equation}
where $g_\sigma$ and $\lambda_\sigma$ are functions of the parameters $\varepsilon$ and $\gamma$.
It is found that, see \Figref{f2}, the dipolar coupling $g_\sigma$ is maximum at $\varepsilon=0$ and decreases with $\gamma$ and $\varepsilon$, while $\lambda_\sigma$ vanishes at large $\varepsilon$ and at $\varepsilon=0$.
The spin-spin interaction is mediated by the resonator, allowing to reach all-to-all connectivity.
At large bias energy $\varepsilon$, the qubits are in memory mode, meaning that they are detuned from the resonator and therefore uncoupled to the other qubits.
At $\varepsilon=0$, the qubit reaches its maximal coupling with the resonator and therefore with the other qubits and $\lambda_\sigma=0$.
We can exhibit the spin-spin interaction using a Schrieffer-Wolff transformation \cite{warren2021robust} to go into the dispersive basis.
In this representation the total Ising Hamiltonian of the processor reads
\begin{equation}\label{sh1}
    H=\sum_k\frac{\omega_k}{2}\sigma_z^{(k)}+\sum_{k\neq j}\frac{J_{kj}}{2}\sigma_x^{(k)}\sigma_x^{(j)}+\sum_k\Omega_k\sigma_x^{(k)},
\end{equation}
where $J_{kj}$ is a function of $\varepsilon$ and $\gamma$.
Hence our initial Hamiltonian is set from the first term in \Eqref{sh1} while the remaining terms correspond to the target Hamiltonian.
In order to tackle the MWIS problem, we have first to find the correspondence between the problem Hamiltonian and the system Hamiltonian and deduce the suited parameters.
\section{Emulator}\label{S3}
For a meaningful emulation of the quantum annealing, we have first to take into consideration the different errors that the processor is subjected to.
There are two types of error that we take into consideration.
First the coherent errors that are linked to the dynamics of the annealing process.
As the parameters of the Hamiltonian evolves in times, the system goes through some anti-crossing at which the condition of adiabaticity may be broken.
These errors are linked to a speed limit of the annealing process as they are typically avoided with operating slow.
Going to fast in the vicinity of an anti-crossing will provoke diabatic transitions.
The second type of errors are the decoherent errors.
They correspond to the system being open and are typically avoided with operating fast.
The interplay between these two types of error gives the optimal operating time of the quantum annealer.
The last challenge that the emulator has to face is its ability to simulate a large enough number of qubits.
To address this challenge we have resorted to a quantum Monte-Carlo simulation to simulate the annealing \cite{rajak2023quantum}.
In the following we will focus on the errors as it is our main contribution to this work.
\subsection{Coherent errors}
To account for the coherent errors, we need to take into consideration the effect of the time evolution of the parameters of the system.
We consider for this work that only the qubits' bias energy $\varepsilon_k$, for $k$ spanning the qubit register, will be modified during the dynamics.
Essentially, the transformation we used to get to the computational basis now becomes time dependent and the transformed Hamiltonian becomes
\begin{equation}
    \tilde{H}(t)=U^\dagger(t)H(t)U(t)+i\dot{U}^\dagger (t)U(t).
\end{equation}
This modifies \Eqref{sh2} by adding an additional Hamiltonian term
\begin{equation}\label{shc}
    H_{\mathrm{c}}=\frac{\Theta}{2}\sigma_y,
\end{equation}
That corresponds to diabatic excitation of the qubit in the computational basis.
$\Theta$ is related to the speed at which the system parameters evolve in time.
The faster the parameters are changed, the bigger $\Theta$.
In the computational basis we still don't have access to the Ising Hamiltonian.
We have to go in the dispersive basis to have access to the Ising Hamiltonian as discussed in the previous section.
This is done using a time dependent Schrieffer-Wolff transformation.
Calling $S(t)$ the generator of the transformation, such that $U_s=\exp(S)$ and $S^\dagger=-S$, our transformation is such that
\begin{equation}
    [H_0(t),S(t)]=-g\left(a+a^\dagger\right)H_D(t)-i\dot{S}(t),
\end{equation}
where
\begin{equation}
    H_0(t)=\frac{\omega_q(t)}{2}\sigma_z+\frac{\Theta(t)}{2}\sigma_y+\omega_ca^\dagger a.
\end{equation}
Defining
\begin{equation}
    S=a(\alpha \sigma_x+\beta\sigma_y+\gamma\sigma_z)-H.c.,
\end{equation}
where $H.c.$ stands for Hermitian conjugate, it amounts to solving the set of differential equations
\begin{equation}
\left\{\begin{aligned}
    &\frac{d}{dt}\left(\alpha e^{-i\omega_ct}\right)=\left(ig_\sigma -\Theta\gamma +\omega_q\beta\right) e^{-i\omega_ct}\\
    &\frac{d}{dt}\left(\beta e^{-i\omega_c t}\right)=-\omega_q\alpha e^{-i\omega_ct}\\
    &\frac{d}{dt}\left(\gamma e^{-i\omega_ct}\right)=\left(\Theta\alpha +i\lambda_\sigma\right) e^{-i\omega_ct},
\end{aligned}\right.
\end{equation}
where all variables are time dependent functions.
This set of equations is solved numerically before the simulation.
The many body Hamiltonian that includes all the qubits in the processor is
\begin{equation}
\begin{aligned}
    H=&\omega_ca^\dagger a+\sum_k\left(\frac{\omega_k}{2}\sigma_z^{(k)}+\frac{\Theta_k}{2}\sigma_y+\right)\\
    &+\left(a+a^\dagger\right)\sum_k\left(g_k\sigma_x^{(k)}+\lambda_k\sigma_z^{(k)}\right)
\end{aligned}
\end{equation}
Finally, defining $S_k=\vec{A}_k\cdot\vec{\sigma}_k$, the natural extension of the Schrieffer-Wolff transformation to all the qubits, where $k$ runs over all the qubits, $\vec{A}_k=(\alpha_k,\beta_k,\gamma_k)$ and $\vec{\sigma}_k=(\sigma_x^{(k)},\sigma_y^{(k)},\sigma_z^{(k)})$ and $U=\exp\left(\sum_kS_k\right)$, we find the Ising Hamiltonian to be \cite{warren2021robust}
\begin{equation}
    H=H_I+H_{c}
\end{equation}
Where $H_I$ is defined in \Eqref{sh1} and $H_c$ in \Eqref{shc}.
As a reminder, all parameter of $H_I$ and $H_c$ are time dependent.
Note that due to the form of $U$,   at first order in $g_k/\Delta_k$, where $\Delta=\omega_k-\omega_c$, there is no mixing of the different qubits in the error term $H_c$, meaning the the errors remain local.
This means that the coherent errors are still controlled from the diabacity of the transitions around anti-crossings for each individual qubits.
\subsection{Incoherent errors}
Incoherent errors are derived from the environment of the qubits.
Indeed the qubit is subjected to several reservoirs, bosonic ones, such as the coupling resonator, the carbon nanotube's phonons, or fermionic ones such as the contact electrodes or the surrounding conductors fluctuations provoking charge noise.
All of these noises are taken into account in the Lindblad formalism, assuming all sources of noise are Markovian.
In our study we focus mainly on the Purcell effect coming from the resonator, the phonons and the charge noise as they are expected to be the most impactful on this technology.
The phonons and photons are both coupled through the same orbital degree of freedom and therefore have very similar effects on the qubit.
We can derive the Lindbladians at all time from the transformation $U$.
For the charge noise, we take into consideration both qubit relaxation and dephasing and once again we update the Lindblad rates for dephasing and relaxation using the transformation $U$.
That way, we obtain time dependent rates in the Lindblad equation that account for the decoherence of the qubits.
The rates keep the same expressions as the usual static ones \cite{warren2021robust}, except that we have to account for the time variations of the parameters $g_\sigma$ and $\lambda_\sigma$ as well as the qubit frequencies.
This decoherence is important to establish the upper limit of the annealing time.
\section{Results}\label{S4}
\subsection{Benchmarking scenarios}

As described in the previous sections, we applied \textit{Callisto} to address the MHT problem. In radar-based target tracking, each time step corresponds to a new set of measurements. Some of these measurements correspond to true targets, while others represent noise or false alarms. Consequently, at every time step, the algorithm must determine which detections are associated with the actual targets and which should be discarded as false positives. Each unique set of associations constitutes one hypothesis. After each update, the number of hypotheses grows exponentially—a manifestation of the combinatorial explosion characteristic to the MHT problem, which is known to be NP-hard. Therefore, at each iteration, the MHT algorithm must evaluate the likelihood of each individual hypothesis and retain only the most probable ones, a process commonly referred to as pruning.

As explained in Section \ref{S1} to address this computational challenge, we formulate the tracking problem as a Maximum Weighted Independent Set (MWIS) problem. In this formulation, each node in the graph represents a hypothesis, with its corresponding likelihood serving as the node weight. The objective is thus to identify the subset of non-conflicting hypotheses with the maximum total weight (as described by Equation \ref{mwis}). We map this graph formulation directly onto a C12 quantum system using the \textit{Callisto} annealer emulator. Through the annealing process, the system efficiently converges to the optimal configuration.

For benchmarking, we follow a similar procedure to that presented in \cite{RoussetRouard2025benchmark}. The results obtained using the Callisto emulator are compared against those from classical algorithms, such as the Two-Stage MaxSAT for the Maximum Weighted Clique (TSM-WC) method introduced in \cite{jiangetal2018}. Our benchmarking is divided into two distinct scenarios:

\begin{enumerate}
    \item \textbf{Single-Step QA}: The emulator is applied only at the time step that corresponds to the highest number of hypotheses for the specific problem.
    \item \textbf{Sequential QA}:The emulator is used at each time step.
\end{enumerate}

In the first scenario, we consider a simplified case involving two objects moving linearly at a constant velocity, with noisy measurement data. We initially solve this problem using the classical TSM-WC approach to identify the time step with the largest number of hypotheses. This is crucial, as the computational cost of TSM-WC increases drastically with the number of hypotheses, making this step a suitable point for demonstrating the advantage of using a quantum annealer.

\begin{figure}
    \centering
    \includegraphics[width=\columnwidth]{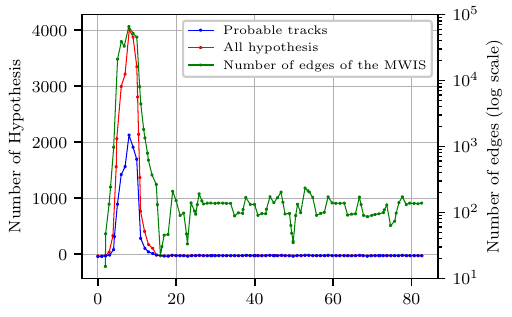}
    \caption{The number of hypothesis as a function of time for a scenario with two objects and noisy measurements, for the $\lambda_{c}=1\times10^{-5}$.}
    \label{fig:bench_sc1_numhyp}
\end{figure}
\begin{figure*}
    \centering
    \includegraphics[width=\textwidth,trim={0 1mm 0 0},clip]{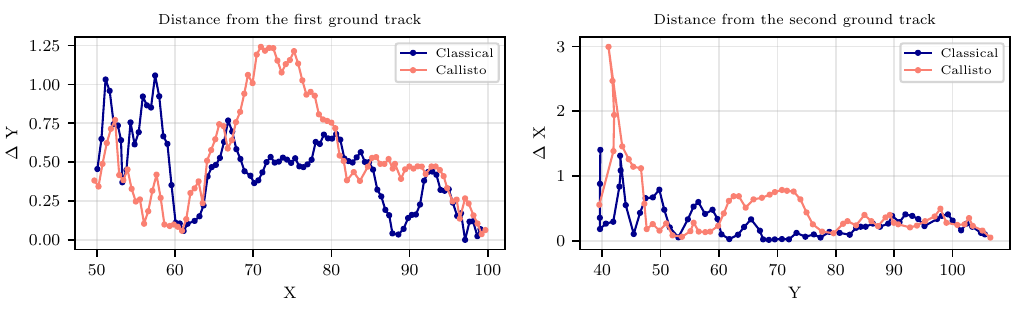}
    \caption{The deviations from the ground tracks obtained  using the TSM-WC classical solver (dark blue) and Callisto (light red) annealer only on the 10th timestep (highest number of hypotheses).  $X$ and $Y$ are the spatial coordinates.}
    \label{fig:becnh_sc1_tracks}
\end{figure*}

Figure \ref{fig:bench_sc1_numhyp} illustrates the number of hypotheses as a function of time, where a sharp increase by nearly three orders of magnitude occurs between time steps 4 and 10. At the 10th time step, we applied the C12 annealer to solve the corresponding MWIS problem. For this specific problem we were using the clutter density ($\lambda_{c}$) value of $1\times10^{-5}$. The tracks obtained are shown in Figure \ref{fig:becnh_sc1_tracks}. This is a parameter that quantifies the expected number of false measurements per unit volume. This value influences the probability that a given measurement is due to clutter rather than a real target (see \cite{RoussetRouard2025benchmark}), therefore it can influence the pruning of the hypotheses. In the MHT framework, both classical and quantum implementations rely on likelihood scores that incorporate this clutter probability when evaluating measurement-to-track associations. A higher clutter density increases the likelihood that measurements are treated as false alarms, which reduces the confidence of certain associations and leads to higher pruning of hypotheses. A lower value of the clutter density favors new measurements originating from real targets, allowing more hypotheses to survive. As a result, the value of $\lambda_c$ directly affects the structure of the hypothesis graph and therefore the MWIS problem solved by classical algorithms or quantum annealer.

\begin{figure}
    \centering
    \includegraphics[width=\columnwidth]{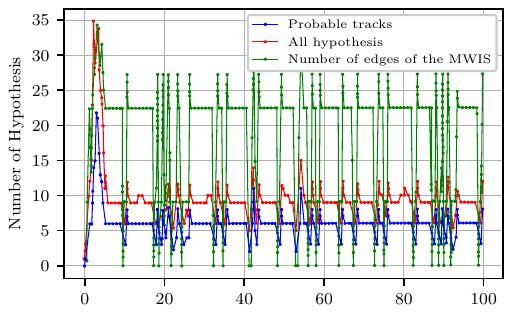}
    \caption{The number of hypotheses as a function of time for a scenario with one object and noisy measurements, for the $\lambda_{c}=1\times10^{-5}$ where Callisto annealer is applied on each time step.}
    \label{fig:bench_sc2_numhyp_small_lambda}
\end{figure}

\begin{figure*}
    \centering
    \includegraphics[width=\textwidth,trim={0 1mm 0 0},clip]{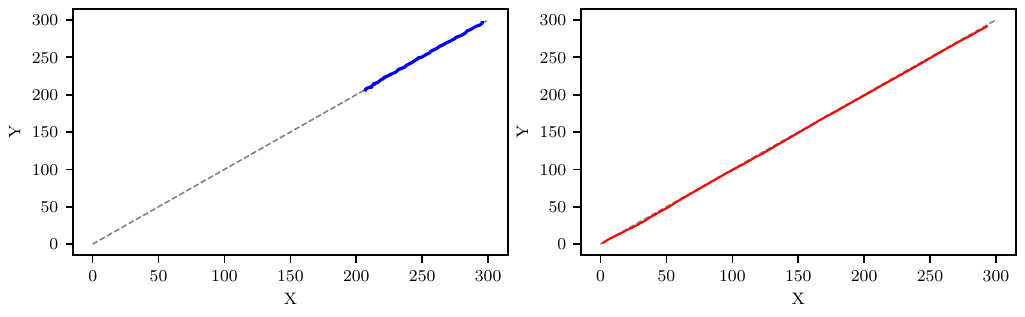}
    \caption{Tracks obtained using the Callisto annealer on each time step for the scenario with one linear track and the $\lambda_{c}=1\times10^{-5}$ (right) and  $\lambda_{c}=5\times10^{-5}$ (left) where Callisto annealer is applied on each time step. The dashed line is the real track and the red line is the recovered track using the algorithm. $X$ and $Y$ are the spatial coordinates.}
    \label{fig:becnh_sc2_tracks_small_lambda}
\end{figure*}

In the second scenario, we perform a single-target tracking experiment with the $\lambda_c$ value of $5\times10^{-5}$, applying the Callisto annealer at every time step (see figure \ref{fig:becnh_sc1_tracks} that demonstrates both the number of hypotheses and the corresponding execution time of the emulator). When the $\lambda_c$ parameter is increased the MHT algorithm exhibits a more conservative behavior, as the higher $\lambda_c$ value increases the probability that individual detection originate from clutter rather than the true target. As a result, some hypotheses can be pruned prematurely which can lead to fragmented trajectory reconstruction, this can be seen in the figure \ref{fig:becnh_sc2_tracks_small_lambda}. To examine the effect of the parameter $\lambda_c$ on the resulting trajectory, we conducted the same single-track experiment using a changed parameter value of $\lambda = 1 \times 10^{-5}$. The increased value of $\lambda_c$ affects the number of generated hypotheses (see Figure \ref{fig:bench_sc2_numhyp_small_lambda}). After applying the Callisto annealer at each time step, the complete trajectory was obtained, as illustrated in Figure \ref{fig:becnh_sc2_tracks_small_lambda}.
\\

\begin{figure}
    \centering
    \includegraphics[width=\columnwidth]{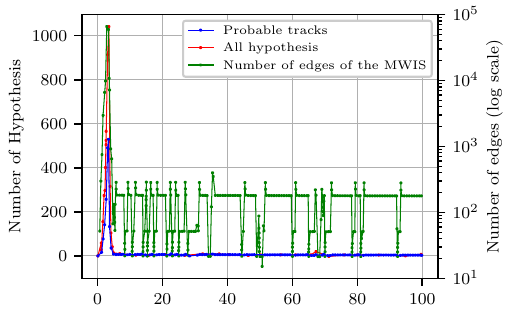}
    \caption{The number of hypotheses as a function of time for a scenario with one object and noisy measurements, for the $\lambda_{c}=5\times10^{-5}$ where Callisto annealer is applied on each time step.}
    \label{fig:bench_sc2_numhyp}
\end{figure}

% %
% For the benchmarking we follow the scenario presented in \cite{RoussetRouard2025benchmark}.
% %
% A linear track is followed over 20 time steps.
% %
% Between step 3 and step 10, the number of false alarms increases by a factor of 100
%
\subsection{Annealing performance}

One of the important aspects of this work is the determination of the execution time of the annealing process using C12 hardware model as it is a key feature that the algorithm is run in real time. The total annealing execution time  consists of the following parts: state preparation, annealing time and read-out time. 
Using the C12 hardware, the state preparation time was measured to be in the range of approximately $5$–$10\;$\si{\milli\second}.
This corresponds to a passive reset of the processor.
A faster reset can be implemented using active feedback protocole \cite{kobayashi2023feedback}.
It consists in a measurement followed with a single gate operation based on the measurement result in order to make sure that the qubits are all in their respective ground states.
Then the preparation would amount roughly to a single readout time for each qubits.
The readout time for spin qubits using cQED readout, is estimated to be $1\;$\si{\micro\second} per measurement \cite{d2019optimal}. When readout can be parallelized among all qubits either using post treatment \cite{ginzel2023simultaneous} or being able to use a single readout resonator per qubits, measuring the complete register only takes the measurement time of a single qubit.
Finally, the annealing time was adjusted to minimize the total computational error, as this parameter can be explicitly controlled within the emulator environment. Based on these evaluations, the optimal annealing time was found to lie in the range of $50-150\;$\si{\micro\second}.
Accounting for the number of measurement shots, we finally get to a time that is on the range of the second when we consider passive reset and on the range of the \SI{50}{\milli\second} when we consider active reset.

\section{Conclusion}

\begin{figure}
    \centering
    \includegraphics[width=\columnwidth]{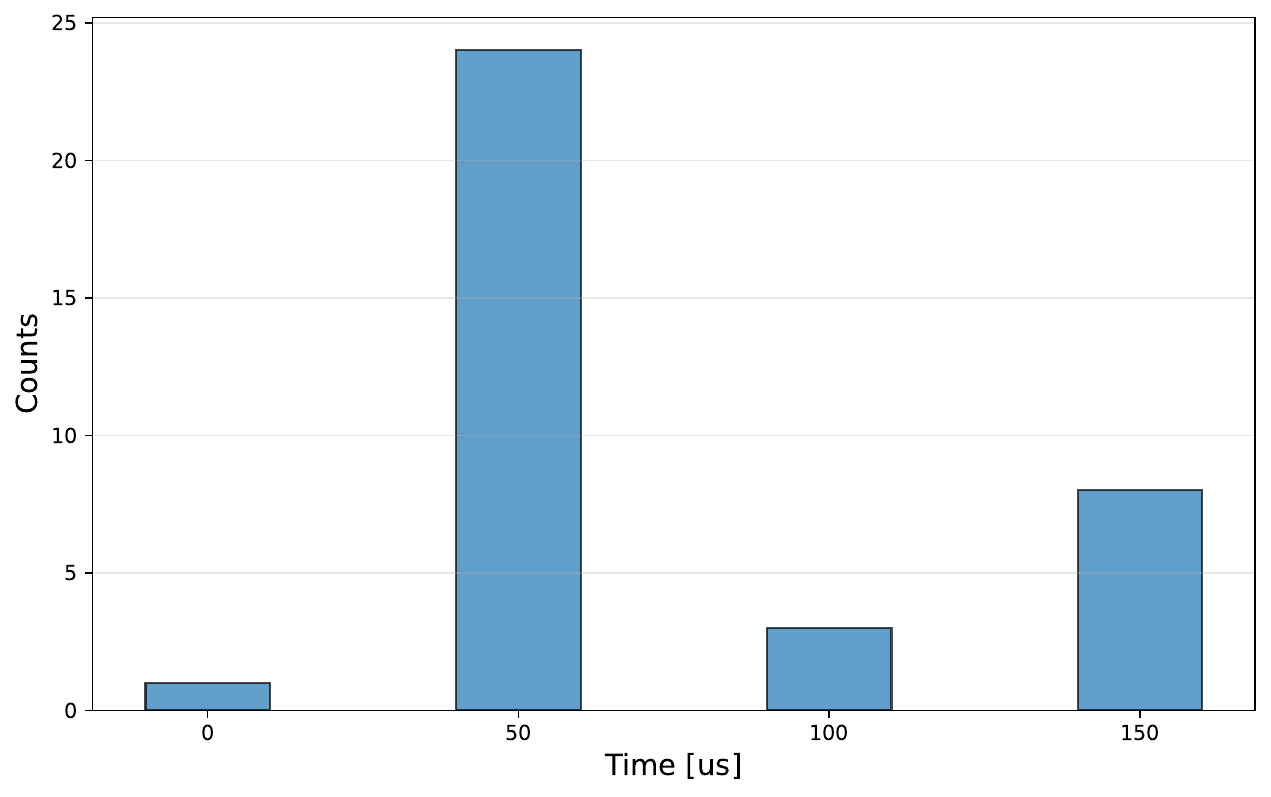}
    \caption{A total operating time histogram for one full run on the Callisto annealer. }
    \label{fig:histogram}
\end{figure}

The ability to electrostatically control quantum dots is often presented as a key advantage in spin qubits to use them either in memory mode or in operating mode \cite{cottet2010spin}.
However it also offers the possibility to use spin qubit processors as annealer instead of gate based digital computers.
Taking advantage of the fast operation of semiconducting spin qubits and the all-to-all connectivity and fast readout offered by cQED architectures, we investigated the usage of cQED spin qubits for a multi-tracking algorithm.
Indeed, the operating time is shown to be optimal in the $50$ to $150$ \si{\micro\second} range when we take into account both the decoherence processes in such technologies (e.g. Purcell effects, charge noise and nuclear spin noise) and the diabatic coherent errors that occur during the annealing procedure.
Again, taking advantage of the cQED approach, we are able to reset and measure the qubit on the micro second time scale.
When the processor allows for parallel qubit measurement and using active qubit reset, this means that the annealing time becomes our limiting factor and we are able to reach a total operating time on the scale $50\;$\si{\milli\second} in the best cases (Figure \ref{fig:histogram}).

On the reverse, if we are only using passive reset, the reset time becomes the limiting factor of the whole process and the total operating time to get to a final result is on the scale of $5\;$\si{second}.
Overall this shows that cQED spin processors are promising in the context of real-time application for problems such as multi-target tracking for radar detection.

\begin{acknowledgments}
    We thank Dr. M. M. Desjardins for fruitful discussions and his careful reading of this manuscript. We thank David Sadek and Frédéric Barbaresco for THALES funding of this study, and Mathias Bossuet and Rami Kassab for THALES LAS/SRA support.
\end{acknowledgments}

\bibliography{bib}

\end{document}